\documentclass[aps,prl,twocolumn]{revtex4}

\usepackage{amsmath}
\usepackage{amssymb}
\usepackage{amsbsy}
\usepackage{makeidx}
\usepackage{epsf}
\usepackage{graphicx}
\usepackage{amsfonts}
\usepackage{subfigure}

\newcommand{\sech}{\operatorname{sech}}
\newcommand{\csch}{\operatorname{csch}}

\begin{document}
\title{Variational approximations to homoclinic snaking}
\author{H. Susanto}
\author{P.C. Matthews}
\affiliation{School of Mathematical Sciences, University of Nottingham, University Park, Nottingham, NG7 2RD, UK}


\begin{abstract}
We investigate the snaking of localised patterns,
seen in numerous physical applications,  
using a variational approximation.
This method naturally introduces the exponentially
small terms responsible for the snaking structure, that are
not accessible via standard multiple-scales asymptotic techniques. 
We obtain the symmetric snaking solutions and the
asymmetric `ladder' states, and also predict the stability of the
localised states.  
The resulting approximate formulas for the width of the snaking region
show good agreement with numerical results. 
\end{abstract}

\maketitle

There has been much recent interest in the phenomenon of spatially
localised patterns, \cite{budd05,burk07,chap09,kozy06}, extending our
understanding of earlier work on this topic \cite{pome86,bens88}.
As discussed in the review article by Dawes \cite{dawe10}, this work
is motivated by wide-ranging applications in many different areas of
physics, including buckling of struts and cylinders
\cite{cham98,hunt00,wade99}, nonlinear optics \cite{chap09},
convection patterns \cite{bens88,nepo94}, gas discharge systems and
granular media. 
Most theoretical work concentrates on the Swift-Hohenberg equation,
which is the simplest model equation that illustrates the effect
\cite{bens88,dawe10,kozy06}.

These localised patterns arise as a result of bistability 
between a uniform state and regular patterned state.
A front linking these two states might be expected to move in one
direction or the other, except at a specific parameter value, referred
to as the Maxwell point. However, due to a pinning effect first
described qualitatively by Pomeau \cite{pome86}, the front can lock to
the pattern, 
resulting in a finite range of parameter values around the Maxwell
point where a stationary front can exist. Combining two such fronts
leads to a pinned localised state.  The pinning range is seen
in numerical simulations \cite{saka96,wood99,burk06,burk07,burk07_2}
and leads to a `snaking' bifurcation diagram in which the control 
parameter oscillates about the Maxwell point as the size 
of the localised pattern increases. 

The localised states can also been seen from the viewpoint of spatial
dynamics as a homoclinic connection from the uniform state to itself,
leading to a geometrical argument for the existence of such states over
a finite range of parameter values \cite{coul00,cham98,hunt00}.

It has been appreciated for some time \cite{pome86} that the pinning effect 
cannot be described by conventional multiple-scales asymptotics, since
this method treats the scale of the pattern and the scale of its
envelope as independent variables, leading to an arbitrary phase in
the envelope function. Hence, the pinning range is `beyond all
orders', or 
is exponentially small in the small parameter corresponding to the
pattern amplitude \cite{bens88}.  A very thorough analysis of the
exponential asymptotics of this problem has recently been carried out by 
Kozyreff and  Chapman \cite{kozy06,chap09}, extending earlier work
\cite{yang97,wade99}.  The calculation of the 
pinning range is extremely complicated and unfortunately requires two
fitting parameters. 

Many previous authors have noted the variational property 
of the Swift-Hohenberg equation \cite{budd05,nepo94,burk07},
but  have not made use of this in regard to the snaking diagram. 
Wadee and Bassom \cite{wade00} did use the variational method, but not
in the snaking regime. 
In this Letter we exploit the variational structure of the
Swift-Hohenberg equation in order to approximate the pinning range. 
An ansatz based on the weakly nonlinear solution is substituted into
the Lagrangian of the system, and then this is minimised over the
unknown parameters.  This approach has a number of advantages: 
the calculations are much less cumbersome than those
involved in the full exponential asymptotics analysis;
exponentially small terms appear naturally through the evaluation of
the Lagrangian integral; there are no unknown constants 
that have to be determined by fitting with numerical results; and the corresponding effective Lagrangian can predict the stability of the states. 
Although the method is widely applicable, for illustrative purposes we
consider the cubic-quintic Swift-Hohenberg equation
\cite{burk07,dawe10,nepo94}. 
The results are then compared with numerics. 

The governing equation is given by
\begin{equation}
\partial_tu=ru-\left( 1+\partial_x^2\right)^2u+b_3u^3-b_5u^5,
\label{gov}
\end{equation}
where $r$ is the control parameter and $b_3$ and $b_5$ are the
coefficients of the cubic and quintic term, respectively. 
The uniform state $u=0$ is unstable for $r>0$.
The interesting case leading to snaking
bifurcations is $b_3>0$, $b_5 > 0$, in which case localised
solutions can exist for $r < 0$. 
It is possible to rescale $u$ to set $b_5=1$ but we will keep $b_5$ as
a parameter for consistency with \cite{budd05,burk07}.

The stationary solutions of (\ref{gov}) can be derived from a Lagrangian
\begin{equation}
\mathcal{L}=\int_{-\infty}^\infty \left( \frac{u_{xx}^2}{2}-{u_x^2}+(1-r)\frac{u^2}2
-\frac{b_3}{4}u^4+\frac{b_5}6u^6 \right)dx,
\label{lag}
\end{equation}
which is finite since we are concerned with localised solutions. 
Furthermore,
\begin{equation}
\frac{d\mathcal{L}}{dt} = - \int_{-\infty}^\infty (\partial_tu)^2 \, dx
\end{equation}
so stable solutions correspond to minima of $\mathcal{L}$
and $\mathcal{L}$ can be interpreted as the total `energy' of a
solution. The $L^2$-norm of the solution is defined as $N=\left(\int_{-\infty}^{\infty}u^2\,dx\right)^{1/2}$.

Motivated by  weakly nonlinear analysis \cite{bens88,burk07,wade99}
we study two forms of localised pattern. 
For small $r$, we write the localised solution of (\ref{gov}) as
\begin{equation}
u=A\sech(Bx)\cos(kx+\varphi).
\label{ua}
\end{equation}

Using, e.g., complex variable techniques, substituting the approximation (\ref{ua}) into the Lagrangian
(\ref{lag}) yields the effective Lagrangian
\begin{eqnarray}
\mathcal{L}_{eff}&=&\frac{A^2}{720B^6}\left[ 2B^5\left(84B^4+120B^2\left(3k^2-1\right)\right.\right.\nonumber\\
&&\left.+5\left\{-9A^2b_3+4A^4b_5+36\left(\left(k^2-1\right)^2-r\right)\right\}\right)\nonumber\\ 
&&\left.+3k\pi \sum_{n=1}^{3}K_n \cos(2n\varphi)\csch\left(\frac{nk\pi}{B}\right) \right],
\label{efl}
\end{eqnarray}
where
\begin{eqnarray}
K_1&=&56B^8+5A^2B^2\left(-8b_3+5A^2b_5\right)k^2+5A^4b_5k^4+\nonumber\\
&&80B^6\left(k^2-1\right)+4B^4\left(-10A^2b_3+5A^4b_5+6k^4-\right.\nonumber\\
&&\left.20k^2+30\left(1-r\right)\right),\nonumber\\
K_2&=&4A^2\left(B^2+4k^2\right)\left(B^2\left(-5b_3+4A^2b_5\right)+4A^2b_5k^2\right),\nonumber\\
K_3&=&A^4b_5\left(4B^4+45B^2k^2+81k^4\right).\nonumber
\end{eqnarray}

Note that the terms involving the phase shift $\varphi$ in (\ref{efl}) have an
exponential factor. Therefore, it is expected that the splitting of
localized solutions with different phase in the limit $r\to0$, i.e.\
$k\to1$ and $B\to0$, is exponentially small, as suggested in \cite{wade99,kozy06,{chap09}}. 
For the ansatz (\ref{ua}) 
\[\int_{-\infty}^{\infty}u^2\,dx=\frac{A^2}{B}\left\{1+\frac{k\pi}{B}\cos(2\varphi)\csch\left(\frac{k\pi}{B}\right)\right\},\]
which is the square of the norm. 

Applying the Euler-Lagrange formulation to the effective Lagrangian
\begin{equation}
\partial_A\mathcal{L}_{eff}=\partial_B\mathcal{L}_{eff}=\partial_k\mathcal{L}_{eff}=\partial_\varphi \mathcal{L}_{eff}=0,
\label{euler}
\end{equation}
gives us a system of nonlinear equations for $A,B,k$, and $\varphi$ that make (\ref{ua}) an approximate solution of (\ref{gov}). Neglecting the exponentially small terms, (\ref{euler}) can be solved perturbatively to yield (up to $\mathcal{O}\left((-r)^{5/2}\right)$)
\begin{eqnarray}
A&=&2\sqrt{\frac{2}{3b_3}}\sqrt{-r}+\left(\frac{128\sqrt{\frac23}b_5}{81b_3^{5/2}}+\frac1{15\sqrt{6b_3}}\right)(-r)^{3/2},\label{Acq}\\
B&=&\frac{\sqrt{-r}}{2}+\left(\frac{7}{120}-\frac{16b_5}{81b_3^2}\right)(-r)^{3/2},\\ 
k&=&1+\frac{r}{8}-\left(\frac{71}{1920}-\frac{8b_5}{81b_3^2}\right)r^2.\label{kcq}
\end{eqnarray}
Note that the leading order expansions above are the same as those obtained
using multiple scale expansions \cite{burk07}. The phase-shift
$\varphi$ at this order is arbitrary, which also agrees with the
multiple scales result. However, taking into account the equation $\partial_\varphi \mathcal{L}_{eff}=0$, in which all the terms are exponentially small, $\varphi$ is a multiple of $\pi/2$, i.e.\ there are two different types of solution, odd or even in $x$. This same conclusion was reported in \cite{wade99} after
a lengthy `beyond all orders' calculation. 

It is expected that the ansatz (\ref{ua}) only resembles exact
solutions of the Swift-Hohenberg equation when $b_3$ is order 1 and
$r$ is small.  Computations of snaking are generally
carried out with neither $r$ nor $b_3$ being small
\cite{budd05,burk07,saka96}. 
For localised states in the snaking regime, the envelope solution has
a plateau, which becomes longer as the norm $N$ increases. 
As the Maxwell point is the center of the snaking, it is reasonable to
construct homoclinic snaking solutions using a special solution at
that point, i.e.\ a front solution.

We approximate the solution in the snaking region by
\begin{equation}
u=\frac{A\cos(kx+\varphi)}{\sqrt{1+e^{B(|x|-L)}}}.
\label{ua2}
\end{equation}
When the modulus sign in the denominator is absent, the solution (\ref{ua2}) forms a front solution at the Maxwell point obtained using a multiple-scales expansion method for small $b_3$ and small $r$ \cite{bens88,nepo94,budd05}. Due to the modulus sign, (\ref{ua2}) patches two fronts of opposite
polarity, leading to a localised state of length $2L$.  Even though
the derivative of this solution is not continuous at $x=0$, it is
smooth enough when $BL \gg 1$, which we assume to be the case (i.e.\
the fronts are well separated). To simplify the algebra we set $k=1$.

Substituting the ansatz (\ref{ua2}) into the Lagrangian (\ref{lag}), the effective Lagrangian can be obtained as
\begin{eqnarray}
\mathcal{L}_{eff}&=&\frac{A^2}{384B}\left(96{B^2}+3{B^4}+72A^2b_3-60A^4b_5\right.\nonumber\\
&&\left.-8LB\left(9A^2b_3+24r-5A^4b_5\right)\right)\nonumber\\
&&+\frac{1}{32B^3}e^{-\frac{2\pi}{B}}\left[10A^6\pi b_5(B^2-2)-16A^4\pi B^2b_3\right.\nonumber\\
&&\left.+A^2B^2\pi(12-B^2-32r)\right]\sin(2L)\cos(2\varphi)\nonumber\\
&&+\frac{1}{32B^2}e^{-\frac{2\pi}{B}}\left[-30\pi A^6b_5+32\pi A^4b_3\right.\nonumber\\
&&\left.+\pi B^2A^2(B^2-4)\right]\cos(2L)\cos(2\varphi)\nonumber\\
&&+\mathcal{O}\left(e^{-\frac{4\pi}{B}}(e^{\pm 2iL},e^{\pm 4iL}),e^{-BL}\right).
\label{efl2}
\end{eqnarray}
To illustrate some of the steps in the calculation, consider the term 
$\int_{-\infty}^\infty u^2 dx$.  This can be written as
\[
\int_{-\infty}^\infty u^2 dx = \frac{A^2}2\int_{-\infty}^\infty
\frac{1+\cos(2x)\cos(2\phi)}{1+e^{B(|x|-L)}}dx,
\]
after using the double-angle formula and the fact that the
envelope function is even. The first term in the integral can be
evaluated directly and gives the answer $A^2 L +\mathcal{O}(e^{-BL})$.
The second term can be found by writing $\cos(2x)=$Re$(\exp(2ix))$ 
and using contour integration. The integrand has simple poles 
at $x=\pm L +i\pi/B, \, \pm L +3i\pi/B, \ldots$, but the first of these
dominates, giving an exponentially small contribution of order  
$e^{-\frac{2\pi}{B}}$. Evaluating the residue in the standard way 
gives the leading-order result
\begin{equation}
\int_{-\infty}^\infty u^2 dx = A^2 L
+2\pi\frac{A^2}{B}e^{-\frac{2\pi}{B}}\sin(2L)\cos(2\phi). \label{u2}
\end{equation} 
The other terms in  $\mathcal{L}$ can be evaluated in a similar way,
using poles of order up to three. 
The formula (\ref{u2}) gives the square of the norm $N$,
indicating that in the snaking region, solutions with a larger norm
correspond to longer plateaus. 


Consider first the terms in (\ref{efl2})
that are not exponentially small, and assume that $B \ll 1$ so that the
$B^4$ term may be neglected.  Making these terms stationary with
respect to $L$, $A$ and $B$ and solving them for $A$, $B$ and $r$
gives $A=A_M$, $a=a_M$, and $r=r_M$, with
\begin{eqnarray}
A_M=3\sqrt{\frac{b_3}{10b_5}},\,a_M=\sqrt{-r_M},\,r_M=-\frac{27b_3^2}{160b_5},
\end{eqnarray}
which are the parameter values of the front at the Maxwell point, in exact agreement with the multiple-scales asymptotic method \cite{bens88,nepo94,budd05}. Note that $\mathcal{L}_{eff}$ depends linearly on $L$, 
but this dependence vanishes at the Maxwell point.
This is to be expected since the Maxwell point is determined by the
condition that the patterned state has the same energy as the zero
state.   

Near the Maxwell point we set 
\[
r=r_M+\delta r .
\]
After making the above simplifications,
\begin{eqnarray}
\mathcal{L}_{eff}&=&\frac{9b_3}{20b_5}a_M+\frac{9 b_3}{1280
  b_5}a_M^3 -\frac{9 b_3}{20 b_5} L \delta r \nonumber\\
&& - e^{-\frac{2\pi}{a_M}}\cos(2\varphi)\left(\frac{\pi a_M}{320 b_3
  b_5}\left[4480b_5 + 57b_3^2\right]\sin(2L) \right. \nonumber\\
&& \left. -\frac{3\pi b_3}{51200 b_5^2}\left[10880b_5 + 81b_3^2\right]\cos(2L)\right) . \label{efl2s}
\end{eqnarray}


In this form, the Lagrangian can be interpreted more easily. 
The first two terms are constant and independent of $L$. 
These arise from the two front regions at the ends of the localised
states. The first of these terms dominates for small $b_3$, being of
order $b_3^2$. Since this term is positive, localised states have a
greater energy than the periodic state or the zero state, so  
energy considerations suggest that the periodic state or the zero
state are `preferred' over localised states. Similarly, 
multi-pulse localised solutions will have a larger value of 
$\mathcal{L}$, in proportion to the number of pulses. 

The third term in (\ref{efl2s}), proportional to $-L\delta r$,
indicates a preference for increasing values of $L$  if $\delta r >
0$, so that the patterned region grows in this case and shrinks if
$\delta r <0$. 

Finally we have the exponentially small terms responsible for the 
snaking. The first of these, proportional to $\sin 2L$, is larger than
the second, for small $b_3$. 

The two remaining undetermined parameters in the ansatz (\ref{ua2}) are
$\varphi$ and $L$.  The variation of the effective Lagrangian with
respect to $L$ and $\varphi$ can be readily calculated from (\ref{efl2s}). 
From $\partial_\varphi \mathcal{L}_{eff}=0$, there are two different
possibilities. The first is that $\sin(2\varphi)=0$,
so as in the previous case, $\varphi$ is a multiple of $\pi/2$. 
Solving the equation $\partial_L\mathcal{L}_{eff}=0$ for $\delta r$ 
at leading order in $b_3$ then gives
\begin{equation}
 \delta r = \pm \frac{560 b_5\pi a_M}{9 b_3^2}e^{-\frac{2\pi}{a_M}}\cos(2L),
\end{equation}
the plus and minus signs corresponding to the odd and even states
respectively. This in turn gives the maximum value of $\delta r$ as
\begin{equation}
\delta r_{m}=\frac{14\pi}{b_3} e^{-\frac{2\pi}{a_M}}\sqrt{\frac{10b_5}{3}}.
\label{drm}
\end{equation}
The width of the snaking region is therefore approximately given by
$2\delta r_{m}$.

The Lagrangian is then an oscillating function of $L$, intertwining between 
the Lagrangian with $\varphi=0$ and $\varphi=\pi/2$. 
There is a periodic sequence of alternating stable and unstable
equilibria, corresponding to minima and maxima of
the Lagrangian respectively. As $\delta r$ is increased, 
all the equilibria disappear in saddle-node
bifurcations at $\delta r=\delta r_{m}$. Near these bifurcations,
the stable state is the one of the pair with the lower value of $L$.

The second possibility arising from $\partial_\varphi
\mathcal{L}_{eff}=0$ is that $\sin(2L)=0$ (considering only the larger
of the two exponentially small terms in (\ref{efl2s})). 
These types of solutions correspond to the `bridge' \cite{wade02} or
`ladder' \cite{burk07} states that link the snaking branches,
and only exist for values of $L$ that are multiples of $\pi/2$. 
For these solutions,  $\partial_L\mathcal{L}_{eff}=0$
determines the value of the phase $\varphi$ by
\begin{equation}
\delta r = \pm\frac{560b_5\pi a_M}{9 b_3^2}e^{-\frac{2\pi}{a_M}}\cos(2\varphi).
\end{equation}

Considering $\mathcal{L}_{eff}$ as a function of the two variables
$\varphi$ and $L$, it is easy to see that the snaking solutions 
are either maxima or minima, while the ladder states are always saddle
points. Therefore, the snaking solutions are either stable, 
or unstable with two positive eigenvalues, and the ladder states 
are always unstable with one positive eigenvalue. These results are in
agreement with numerical simulations (see \cite{burk07}). 



\begin{figure}[tbhp]
\centering
\includegraphics[width=7cm,clip=]{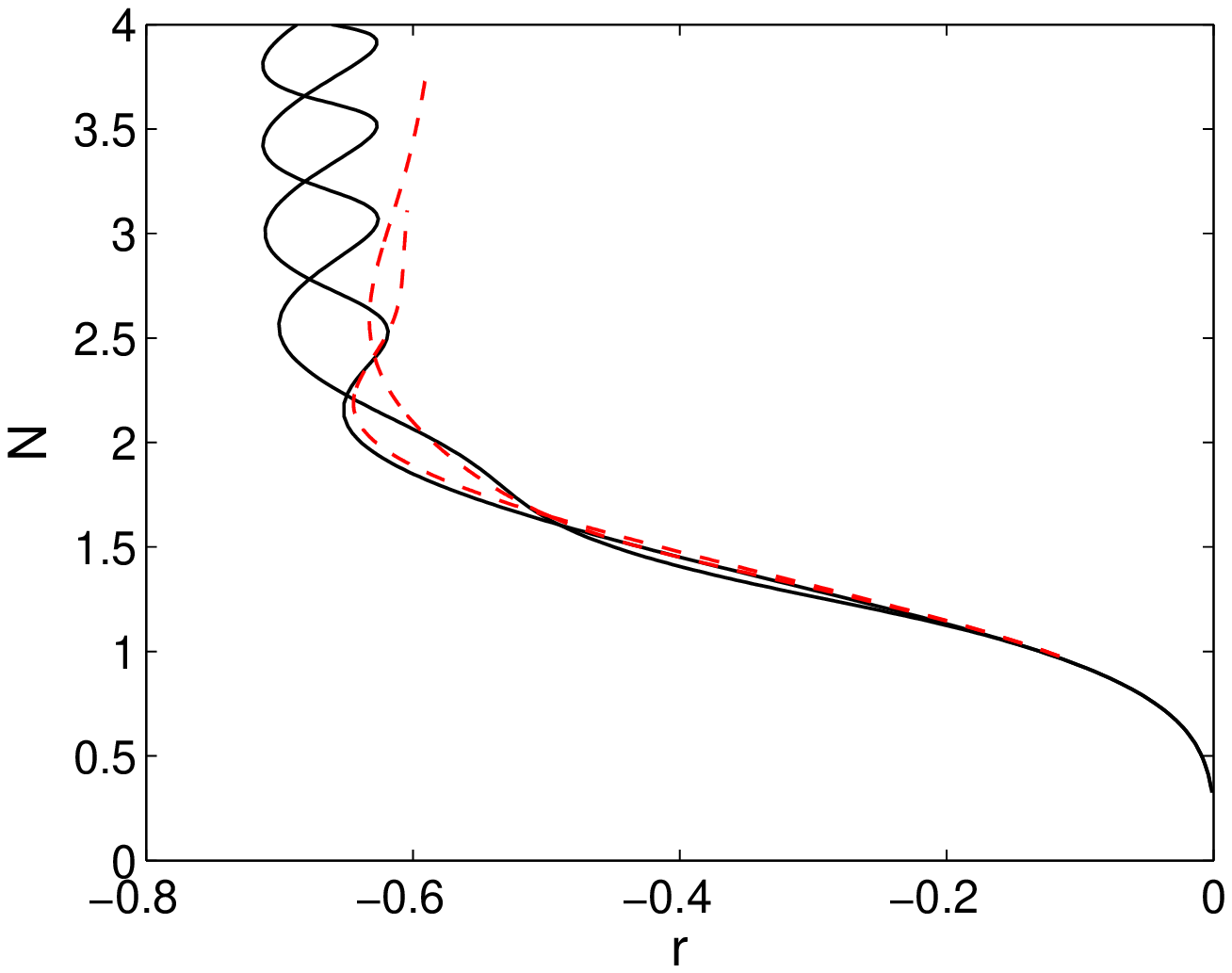} \\ 
\includegraphics[width=7cm,clip=]{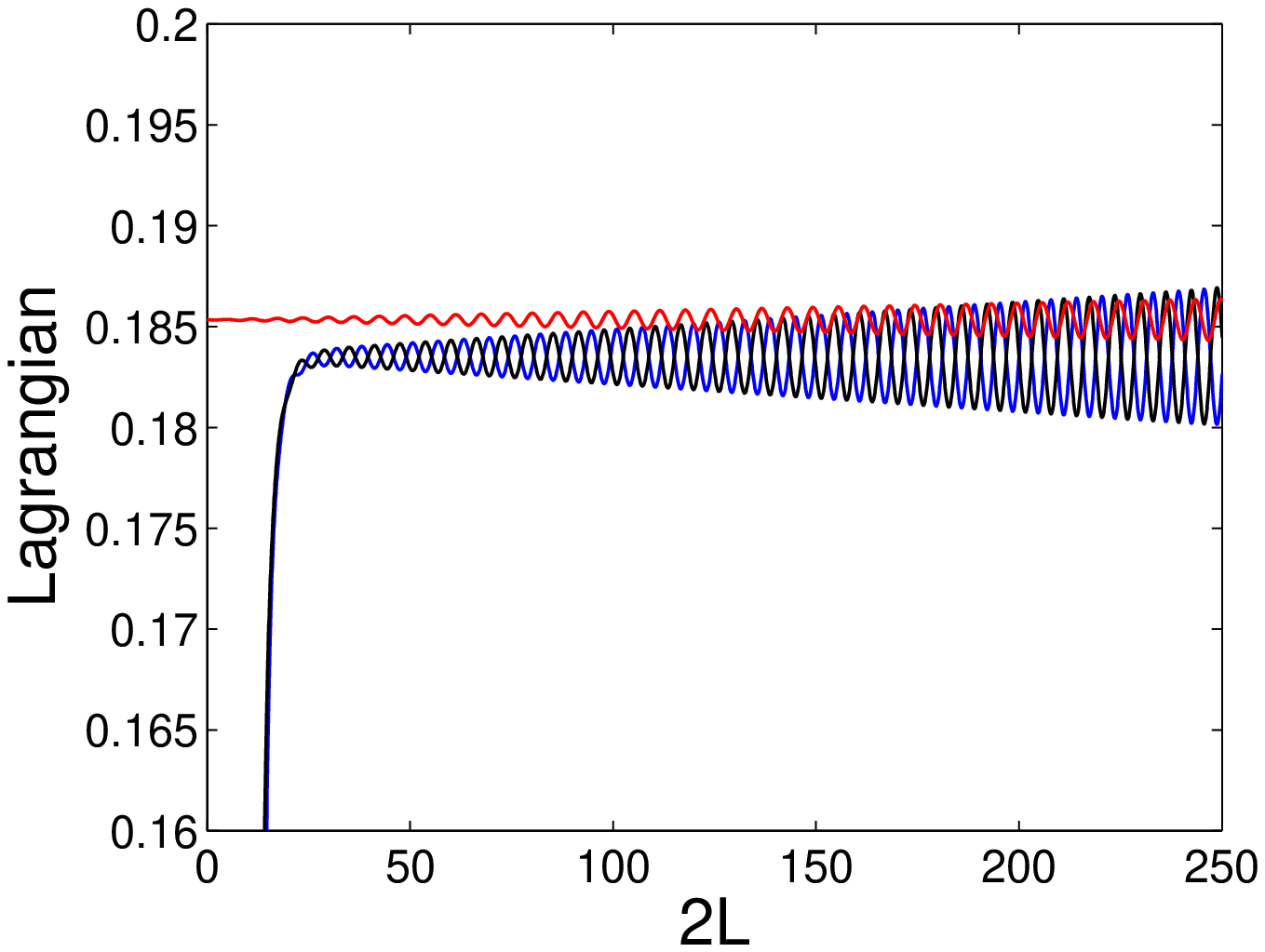} \\
\includegraphics[width=7cm,clip=]{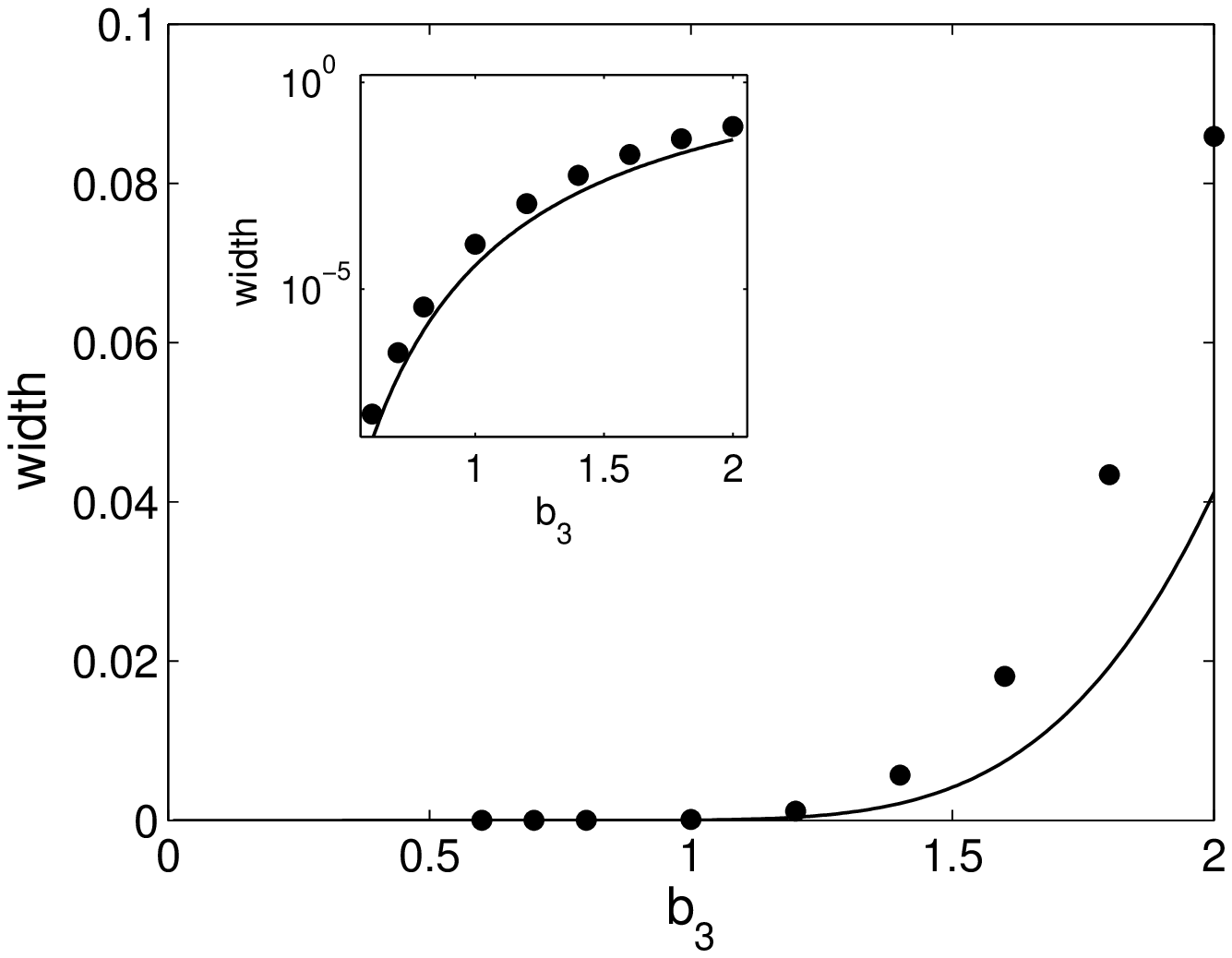}
\caption{(Top) The bifurcation diagram obtained numerically and our
  approximation obtained from solving (\ref{euler}) for
  $b_3=2$. (Middle) The numerically computed Lagrangian (\ref{lag}) as
  a function of the length of the plateau $2L$, for $b_3=1$. The dashed line is our approximation calculated from the effective Lagrangian (\ref{efl2}). (Bottom) The width of the snaking region as a function of $b_3$. Filled circles are numerical and solid lines are analytical, i.e.\ $2\delta r_m$ (\ref{drm}). The inset shows the same comparison in a log scale. In all the figures, $b_5=1$.}\label{figCQ}
\end{figure}

We have found steady localized states of (\ref{gov}) numerically,
using a pseudo-arclength continuation method with periodic
boundary conditions, implemented with a Fourier spectral
discretisation. A summary of the results is shown in Fig.\ 
\ref{figCQ}.

The top panel shows the bifurcation diagram showing two branches of
localized solutions for $b_3=2$ and $b_5=1$. In the same panel, shown
in dashed lines are our analytical results (\ref{euler}), showing
that the variational calculation approximates the numerics
better for relatively small $|r|$. 


In the middle panel, we plot the Lagrangian (\ref{lag}) for $b_3=1$ as a function of the length of the solution's plateau $2L$, which is calculated numerically as $2L\approx\frac2{u_M^2}\int u^2\,dx,$ where $u_M=\max\{u\}$ and the integration is a definite integration over the computational domain. Plotted in the same panel is our effective Lagrangian
(\ref{efl2}). There is good agreement for the average numerical value of the
Lagrangian and the qualitative nature of the oscillations,
but the variational approximation underestimates the amplitude of the oscillations.
The amplitude of the oscillations 
increases with $L$ since from (\ref{efl2s}) with $\delta r \propto \cos(2L)$,
there are oscillations of the form  $L  \cos(2L)$.

In the bottom panel we show the width of the snaking region as a function of $b_3$ numerically and analytically, where our approximation is in fairly good agreement.

To conclude, we have used variational methods to study the
snaking behaviour of localised patterns in the Swift--Hohenberg
equation. The approach 
has several advantages:
the  exponentially small terms responsible for the phase locking arise
naturally in the Lagrangian, giving a simple formula  (\ref{efl2s})
from which the snake and ladder localised states can
easily be found, along with their stability. 
We have concentrated on a simple model equation
in the  small-parameter regime, but  
the method is very widely applicable and can be expected to open up 
a new avenue of research into this challenging field.


\begin{thebibliography}{99}

\bibitem{bens88} D. Bensimon, B. I. Shraiman and V. Croquette,
  Phys. Rev. A 38, 5461 (1988).
\bibitem{budd05} C. J. Budd and R. Kuske, Physica D 208, 73 (2005).
\bibitem{burk06} J. Burke and E. Knobloch, Phys. Rev. E 73, 056211 (2006).
\bibitem{burk07} J. Burke and E. Knobloch, Phys. Lett. A 360, 681 (2007).
\bibitem{burk07_2} J. Burke and E. Knobloch, Chaos 17, 037102 (2007).
\bibitem{cham98} A. R. Champneys, Physica D 112, 158 (1998).
\bibitem{chap09} S. J. Chapman and G. Kozyreff, Physica D 238, 319-354 (2009).
\bibitem{coul00} P. Coullet, C. Riera, and C. Tresser, Phys. Rev. Lett. 84, 3069 (2000).
\bibitem{dawe10} J.H.P. Dawes, Phil Trans. R. Soc. 368, 3519 (2010).
\bibitem{hunt00} G. W. Hunt et al. Nonlinear Dynamics 21, 3 (2000).
\bibitem{kozy06} G. Kozyreff and S. J. Chapman, Phys. Rev. Lett. 97, 044502 (2006).
\bibitem{nepo94} A. A. Nepomnashchy, M. I. Tribelsky and M. G. Velarde,
  Phys. Rev. E 50, 1194 (1994).
\bibitem{pome86} Y. Pomeau, Physica D 23, 3 (1986).
\bibitem{saka96} H. Sakaguchi and H.R. Brand, Physica D 97, 274 (1996).
\bibitem{wade99} M. K. Wadee and A. P. Bassom, Proc. R. Soc. Lond. A
  455, 2351 (1999).
\bibitem{wade00} M. K. Wadee and A. P. Bassom,  J. Eng. Math. 38, 77 (2000).
 \bibitem{wade02} M. K. Wadee, C. D. Coman, and A. P. Bassom,
  Physica D 163, 26 (2002). 
\bibitem{wood99} P. D. Woods and A. R. Champneys, Physica D 129, 147 (1999).
\bibitem{yang97} T.-S. Yang and T. R. Akylas, J. Fluid Mech. 330, 215 (1997). 
\end{thebibliography}
\end{document}